\documentclass[apj]{emulateapj}
\slugcomment{{\sc Accepted to ApJ:} October 11, 2006}
\shorttitle{HVC Distances Through Morphology}
\newcommand{\beq}{\begin{equation}}
\newcommand{\eeq}{\end{equation}}
\newcommand{\hyd}	{{\rm H}}
\newcommand{\vobs}{v_{\rm obs}}
\newcommand{\citei}[1]{\citeauthor{#1} \citeyear{#1}}
\newcommand{\unit}[1]{\textrm{ #1}}

\begin{document}

\title{Reconstructing Deconstruction: High-Velocity Cloud Distance Through Disruption Morphology}

\author{J.~E.~G.~Peek\altaffilmark{1}, M.~E.~Putman\altaffilmark{2}, Christopher F.~McKee\altaffilmark{3}, Carl Heiles\altaffilmark{1}, Sne\v zana Stanimirovi\'c\altaffilmark{4}}

\altaffiltext{1}{Department of Astronomy, University of California, Berkeley, CA 94720}
\altaffiltext{2}{Department of Astronomy, University of Michigan, 500 Church Street, 
Ann Arbor, MI 48109}
\altaffiltext{3}{Departments of Physics and Astronomy, University of California, Berkeley, CA 94720}
\altaffiltext{4}{Department of Astronomy, University of Wisconsin, 475 N. Charter Street, Madison, WI 53706}
\begin{abstract}

We present Arecibo L-band Feed Array 21-cm observations of a sub-complex of HVCs at the tip of the Anti-Center Complex. These observations show morphological details that point to interaction with the ambient halo medium and differential drag within the cloud sub-complex. We develop a new technique for measuring cloud distances, which relies upon these observed morphological and kinematic characteristics, and show that it is consistent with H$\alpha$ distances. These results are consistent with distances to HVCs and halo densities derived from models in which HVCs are formed from cooling halo gas.

\end{abstract}


\section{Introduction}

High-velocity clouds (HVCs) have been detected all over the Galactic sky in the 21-cm hyperfine transition of neutral hydrogen (for an in-depth review see the book edited by  \citei{WWSB2004}). HVCs are clouds with typical velocities between $-300$ and 300 km s$^{-1}$ in the Galactic Standard of Rest (GSR) that cannot be explained by Galactic rotation. Distances to HVCs, and therefore their physical scope, are uncertain and contentious, but recently constraints have been put on the distances to some clouds. The rare direct absorption line measurements of halo stars typically provide distance lower limits of $2$ kpc and include a few upper limits of $10$ kpc (e.g. \citei{Thom2006}, \citei{Wakker2001} for a review). Successful measurements of H$\alpha$ flux from HVCs, which indicate the ionization of the HVCs by escaping Galactic radiation, suggest that many HVCs are within 40 kpc, though the inferred distances have large uncertainties  (\citei{B-HP2001}, \citei{Putman2003}). It is crucial that we understand the true physical size of these HVCs; they have been invoked to provide the accreting material that fuels star formation in our Galaxy (e.g. \citei{RP2000}) and may indeed be a signature of the primary way that baryons cool to form galaxies like the Milky Way. HVCs may also lend insight into the Galactic formation process as a tracer of the evolution of the Galactic halo (e.g. \citei{MB2004}). 

HVCs are thought to move through the Galactic halo. Observations of O\textsc{vi} absorption on sightlines to HVCs have shown that there is collisional ionization in the HVCs, presumably from interaction with the ambient medium \citep{Sembach2003}. Also, many HVCs are seen to have `head-tail' structures, wherein the cloud has a colder high column density core, surrounded by a warmer HI shell which trails off in one direction from the cloud with an associated velocity gradient, as if the cloud were being ablated by the medium (e.g. \citei{BKKM2000}). \citet{Wolfire1995} (hereafter, W95) showed that the two-phase structure of some HVCs implies a significant pressure in the ionized halo that declines with height above the Galactic disk. Subsequently, \citet{SMW2002} showed that this two-phase structure implies that HVCs reside in the Galactic halo and not the Local Group halo. 

We have obtained an image in the 21-cm line of a small sub-complex of high-velocity clouds (HVCs) located at the tip of  the Anti-Center Complex with the Arecibo telescope \footnote{The Arecibo Observatory is part of the National Astronomy and Ionosphere Center, which is operated by Cornell University under a cooperative agreement with the National Science Foundation.}. Arecibo paired with the Arecibo L-Band Feed Array (ALFA), a large filled-aperture telescope with 7-fold multiplexing, affords extremely high sensitivity and dramatic resolution \citep{Stanimirovic2006}. This combination allows us to study the morphology of large clouds and their interfaces with ionized ISM phases in fine detail as never before. The data suggest interaction between the halo gaseous medium and the HVC itself. We claim that by modeling this interaction we can gain insight into the physical attributes of the cloud and the medium through which it moves. Though drag has been studied before in the context of HVCs (e.g. \citei{BD1997}), it is our intent to model \emph{differential} drag within this cloud group, which is immediately evident in the morphology and velocity structure. In particular, these calculations can give us another handle on the distance to the cloud in question, as well as the density of the ambient medium.

\section{Observations}

We observed a region of sky 16x7 square degrees centered on $\alpha$ =  ${\rm 2^{h}15^{m}}$, $\delta$ = ${\rm +9^\circ30^m}$  in May and June of 2005 with Arecibo's ALFA receiver and the GALSPECT spectrometer. ALFA is a new 7-element focal-plane array primarily designed for 21-cm observations. GALSPECT is a special-purpose spectrometer for Galactic science with ALFA. GALSPECT has a spectral resolution of 0.18 km s$^{-1}$, and a fixed bandwidth of 1380 km s$^{-1}$. Each of the 7 beams of ALFA has a 3.35 arcminute half-power beam width with a beam ellipticity of 0.2. The region was initially selected for significant structures in the local, low velocity 21-cm ISM, and the subsequent imaging of HVC 160.7-44.8-333 (hereafter HVC-A), HVC 162.8-46.3-303 (HVC-B) and HVC 156.3-45.1-304 (HVC-C) was purely coincidental. The region was observed in a `basketweave' or meridian-nodding mode, interlacing scans from day to day, similar to observations detailed in \citet{Stanimirovic2006}.

\section{Data Reduction and Analysis}

The raw spectra were calibrated for the IF bandpass using a frequency switching technique that employs a least-squares solution to distinguish the RF spectrum from the IF bandpass \citep{heiles05}. 
The spectra were then cleaned for baseline ripple caused by reflections in the optical fiber between the receivers and the spectrometer. After Doppler-correcting the data to the LSR frame, the data were then corrected for fixed-pattern ripple \citep{heiles05b}. 
 Due to reflections in the Gregorian dome of the Arecibo telescope, ripples in the spectra of order 0.2 K amplitude can occur with a typical `wavelength' of 300 km s$^{-1}$. This ripple is rather constant over the course of a few hours in this observing mode for a single beam, but varies strongly amongst beams. This ripple was mostly eliminated by fitting out residual ripple as compared to the overall average of each day's observations. Each scan's data were then compared with all other scans' data at the sky position where they cross to determine the relative amplitude gains of each beam on each day. The data were re-calibrated for these relative gains, re-reduced for fixed-pattern ripple and then gridded in a 2' x 2' map (see \citet{Stanimirovic2006} for details). The gridded data were then scaled to the equivalent region in the Leiden-Dwingeloo Survey (LDS) for a single, overall gain calibration \citep{Hartmann1997}. 

Once this final data cube was attained, the HVC was fitted with a single Gaussian at each pixel to determine basic characteristics of the cloud. In our data cube, multiple velocity components do not occur in the same pixel, and the single Gaussian fit accurately reports the fluxes, velocities and line widths in the data cube.

\section{The HVCs}

\begin{figure}
\includegraphics[scale=.65]{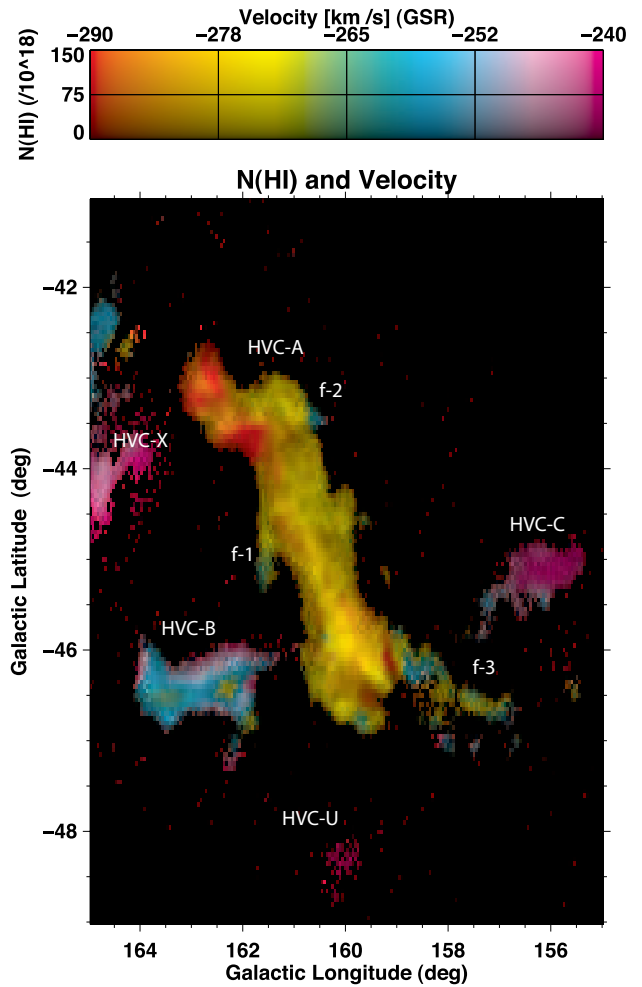}
\caption{The image shows the HI column density and the central velocity along the line of sight, derived by modeling each HI spectrum with a single Gaussian function. Color represents the center velocity of the Gaussian in the GSR frame and brightness represents column density, as shown in the color bar (top). Pictured in full are three clouds, HVC 160.7-44.8-333 (HVC-A), HVC 156.3-45.1-304 (HVC-C) and HVC 162.8-46.3-303 (HVC-B), and the incomplete edge of HVC 166.9-43.2-275 (HVC-X). Also pictured is the very faint  HVC 160.0-48.4-286 (HVC-U). Features are also labeled that are used in our modeling.}
\label{gfit}
\end{figure}

The HVCs imaged are some of the highest Galactic standard of rest (GSR) velocity clouds known, at $-280$ km s$^{-1}$, and can be referred to as very high velocity clouds (VHVCs). \citet{putman2002} showed that the distribution of HVCs in the HIPASS data set (all $\delta< 2^\circ$, resolution of 15.5') can be modeled with a Gaussian distribution centered at $-38$ km s$^{-1}$ GSR with a dispersion of 115 km s$^{-1}$. The updated \citet{WVW1991} catalog (all sky, resolution $<$ 30') has a mean of $-45$ km s$^{-1}$ with a dispersion of 104 km s$^{-1}$.The catalog has been updated by including clouds from the catalog of HVC components by \citet{Morras2000} for $\delta < 23^\circ$ and includes clouds with $|$VLSR$|$ $>$ 90 km s$^{-1}$ (see \citet{Wakker2004} for more information on the catalog properties). It is clear that an HVC with a GSR velocity of $-280$ km s$^{-1}$ is at an extreme of the velocity distribution, though not inconsistent with these samples. We observed 3 large, separated clouds, HVC-A, HVC-B and HVC-C that have all been previously cataloged (see Figure \ref{gfit}). These three clouds have a mean full-width at half-maximum (FWHM) for a single gaussian fit of 28 km s$^{-1}$, with a dispersion of 8 km s$^{-1}$. This is a typical FWHM value for HVCs, in agreement with the 34 km s$^{-1}$ median value from \citei{deheij2002}. We do not have a complete image of the fourth large cloud that is visible in the LDS map of this sub-complex, HVC 166.9-43.2-275 (HVC-X), so we will leave it out of the analysis. Also pictured is the previously unknown HVC 160.0-48.4-286 (hereafter HVC-U). HVC-U is angularly compact ($\sim 30'$) and in any given pixel approaches our detection limit ($\sim 10^{19} {\rm cm^{-2}}$).  Note that this is an example of the detection of a new HVC with ALFA, and confirms the claim that ALFA will detect new low column density clouds in the Galactic halo \footnote{see the GALFA whitepaper, \texttt{http:$//$www.naic.edu$/$alfa$/$galfa$/$docs$/$galfa$\_$white$\_$sept25.pdf}}. 

\citet{WVW2001} have inferred a distance to the observed sub-complex of $13^{+7}_{-5}$ kpc. This distance was measured using the H$\alpha$ flux, under the assumption that all of this flux comes from ionization of HI due to the Galactic interstellar radiation field (ISRF) and subsequent radiative recombination. It is important to note, however, that the model for the UV flux in the halo does not take into account possible patchiness in this radiation field that could arise from variability in the underlying structure of sources of UV radiation (O and B stars), self-shadowing of HVCs to reduce H$\alpha$ flux or possible effects of collisional ioniozation (see \citei{B-HP2001} for an in-depth discussion). This model also fails to predict the observed H$\alpha$ brightness of the Magellanic Stream. In light of these possible sources of error, an independent confirmation of this distance result would significantly bolster the finding.

Morphologically, HVC-A is distinctly elongated, with an aspect ratio of around 6. It also has a few small `arms' (labeled f-1, f-2, and f-3 in Figure \ref{gfit}) that are folded back along the direction of elongation of the cloud. These `arms' typically seem to show some amount of slowing as compared to the bulk of the cloud, as if they are being swept back by the motion of the cloud through the surrounding medium. This velocity gradient may be related to the action that generates `head-tail' morphology observed in other HVCs, which may be caused by ram pressure and viscous stripping \citep{QM2001}. Each of these `arms' points  within the same 45$^\circ$ cone in the plane of the sky, strongly suggesting that the velocity component of HVC-A in the plane of the sky is toward the upper left of the figure, generally along the length of HVC-A. This is consistent with the fact that the surface of HVC-A is much smoother on this `upwind' side, as if sculpted like a raindrop (see models in \citei{QM2001}). We extrapolate further that the smaller three clouds were once part of HVC-A and that the action that is affecting the `arms' on HVC-A has a similar effect upon HVC-B, HVC-C and HVC-U; all of these clouds have relatively similar velocities (see Table \ref{data}) and are `downwind' of HVC-A. We also detect a faint bridge from f-1 to HVC-B, which lends credence to these claims. Under these assumptions, we can learn much about the clouds' physical attributes through the study of gas drag. 

\section{HVC Drag} \label{hvcdrag}

To make headway in understanding the interaction of this cloud with its environment we need to make a number of simplifying assumptions. It is not the authors' intent that these assumptions be exactly correct, but rather that they are plausible; in addition, they give us an idea of the accuracy of the result and the ways in which it is susceptible to systematic errors. Before we begin, let us define a `feature' as any part of the cloud whose movement with respect to HVC-A we wish to model, whether it be a swept-back arm of the main body (i.e. f-1, f-2 and f-3 in Figure \ref{gfit}) or a distinct cloud, such as HVC-B, HVC-C and HVC-U. Three key assumptions regarding the interaction of HVC features with the ambient halo gas to keep in mind in the following analysis are (1) that the surrounding halo gas that generates the drag on these HVCs is static enough that its motion is negligible, (2) that the variation in the direction of the 'apparent wind' that the features experience is small enough to be ignored, and (3) that the displacement of a feature from its origin is governed by the effect of drag.

\citet{BD1997} (hereafter BD97) proposed that the distances to HVCs could be determined by making the assumption that they were at terminal velocity. A cloud moving at terminal velocity in a gravitational field $g$ can be described by a simple force-balance equation:
\begin{equation} \label{tervel}
M_c g = \frac{1}{2}C_D \rho_h {v_c}^2  A,
\end{equation}
where $v_c $ is the velocity of the cloud through the medium, $\rho_h$ is the density of the medium it is moving through, and $A$ is the area that the feature presents to the medium perpendicular to the direction of motion (BD97). $M_c$ is the mass of the feature and g is the acceleration felt by the cloud due to Galactic gravity. $C_D$ here is the `drag coefficient', a way of parameterizing the dependence of drag forces on shape - aerodynamic shapes can have $C_D <$ 0.1 and bluff shapes can have $C_D >$ 2. The basic premise of BD97 is that given (1) a fiducial model of the gas-density of the halo, (2) the assumption that all HVCs are at their terminal velocities, and (3) that their space-velocities are roughly equivalent to their observed velocities, we can find distances to these HVCs. Applying this reasoning and the ISM models used in BD97 to the 3 clouds of interest here, would put the sub-complex at a distance of $\sim$4 kpc (see Table \ref{param} and Figure \ref{rhoD}). Alternatively, if we assume the ${\rm H\alpha}$-measured distance of 13 kpc, the predicted density of the halo medium is an order of magnitude larger than the model assumed in BD97. It is possible to reconcile the terminal-velocity picture with the specific information pertaining to this HVC only by strong fine-tuning of the relevant parameters (e.g. setting the space velocity of the clouds to the Galactic escape speed \emph{and} setting $C_D = 2$). 

In the following analysis, we do not assume that these HVCs are at their terminal velocities, though they experience some drag from the Galactic halo. The equation of motion for a cloud in the Galactic halo that is subject to drag by the gaseous halo is
\beq
M_c a=\frac 12 C_D\rho_h {v_c}^2 A+M_cg,
\eeq
where $a$ is the acceleration.
We have assumed that magnetic fields are weak and
do not significantly affect the motions of the clouds. Two clouds at approximately the same location suffer
a differential acceleration
\beq \label{delta}
\Delta a=\frac 12 C_D\rho_h\Delta\left(\frac{v^2}{\Sigma_c}\right)
\simeq \frac 12 C_D\rho_h v_c^2\Delta\left(\frac{1}{\Sigma_c}\right),
\eeq
where $\Sigma_c\equiv M_c/A$ is the surface density of
the cloud, and where we have used the fact that the clouds
have similar velocities. Clouds suffering a differential
acceleration $\Delta a$ will develop a differential velocity
$\Delta v$ and a separation $s$ related by
\beq \label{kine}
s=\frac{\Delta v_c^2}{2\Delta a}.
\eeq
This assumes that the clouds started at the same velocity before separation. It follows from Eqns. \ref{delta} and \ref{kine} that the number density in the halo
(measured in hydrogen atoms cm$^{-3}$) is
\beq
n_h=\frac{1}{C_D}\left(\frac{\Delta v_c}{v_c}
	  \right)^2\frac{1}{s\Delta N_{\hyd}^{-1}},
\label{eq:nh1}
\eeq
where $N_\hyd$ is the cloud column in hydrogen atoms cm$^{-2}$.

      To put this equation in terms of observables, we
must introduce the angle formed by the overall cloud complex velocity vector
with respect to the line of sight. We call this angle $\phi$.
The observed velocity is then
\beq
|\vobs|=v_c\cos\phi,
\label{eq:vc}
\eeq
where $\vobs$ is negative in this case. The spatial separation
of the clouds $s$ is given in terms of their observed angular
separation $\Theta$ by
\beq
\Theta D=s\sin\phi,
\eeq
where $D$ is the distance to the clouds.
Equation ({\ref{eq:nh1}) then becomes
\beq
\Xi\equiv n_h D= \frac{1}{C_D}\left(\frac{\Delta \vobs}{\vobs}
	       \right)^2\frac{\sin\phi}{\Theta\Delta N_\hyd^{-1}}.
\eeq

In the definition of $\Xi$ the only two variables not measured by observation are $C_D$ and $\phi$. We make the assumption of $C_D$ = 1 (as in BD97) as a fiducial estimate. To estimate $\phi$ we take advantage of the fact that the HVC complex under investigation is at a very high velocity and so will typically have a smaller $\phi$ than the general HVC population; were the $\phi$ to be large the true velocity would be anomalously large (see eqn \ref{eq:vc}) and we would expect to see other HVCs with much larger observed velocities. To quantify this assertion we constructed a mock population of HVCs that correspond to the velocity distribution observed in the updated \citet{WVW1991} catalog. We give these HVCs a Maxwellian distribution of true velocities so that when a na\"ive random distribution of $\phi$ is taken into account the mock velocity distribution reflects the broadly Gaussian catalog velocity distribution. In the catalog, the HVCs in question are on the tail of the distribution, at $\sigma = 2.3$. We investigate this same region in the mock population, and find that clouds at this same velocity have $\phi = 25 \pm 10$ degrees. We take 25 degrees as our fiducial value for $\phi$, and note that a systematic error in $\phi$ can result in errors in $\Xi$ of $\simeq 40\%$.

To reproduce the Gaussian distribution of observed velocities, this analysis makes the assumption that the true velocity unit vectors can point anywhere in space with the bias towards net infall.  While this velocity distribution is what one would expect from models like those described in \citet{MB2004}, these models are still far from general consensus. For rougher limits that do not depend upon models we can assume that the true velocity of this cloud does not exceed the highest velocity ever observed in an HVC in the halo, currently -450 km/s \citep{deheij2002}. This gives us a maximum $\phi$ of 51$^\circ$. For $\phi$ less than this maximum value, there is only an 8\% chance that $\phi < 15^\circ$, our minimum value from the mock distribution. Even with this extreme range of $\phi$, $\Xi$ only varies by a factor of 2.

To do our modeling, we apply this equation to six distinct features that we believe may have been pulled off of HVC-A (see Figure \ref{gfit}). Three of these features, f-1, f-2 and f-3, are small `arms' connected to this main body, and certainly have the swept-back appearance that led us to this reasoning in the first place. The other three regions, HVC-B, HVC-C and HVC-U, are the three other distinct clouds in the map assumed to have been once connected to HVC-A.  We neglect modeling HVC-X as it is not completely imaged. We do not attempt to model other features as difficulty in the evaluation of model parameters ($\Theta$, $\Delta \vobs$ or $N_\hyd$) precludes their accurate analysis. In particular, features with confusing morphology have no reliable value for $\Theta$.

To determine a value for $\Delta N_\hyd^{-1}$ we first note that our measurement is of the neutral hydrogen column density, rather than the total proton column density, so we use a fiducial correction factor 4/3 to account for a small fraction of ionized gas \citep{Tufte2004}.  We then compare the average column density in each feature to the average column density in the `head' of HVC-A, located at $l$ = 162.7, $b$ = $-43.2$. The column density of the front of HVC-A, which is presumed to generate the drag, is the value of interest. Over the disruption timescale, 10 Myr, the cloud travels only $\sim$5 kpc, so our implicit assumption that the density of the ambient medium is constant over the disruption time is not wholly unreasonable.

The errors in the analysis are dominated by systematics such as our measurement of the distance between the features and the main cloud and the unknown 3D orientation and movement of HVC-A and the other features. Since the values of $\Xi$ we predict vary with a fractional standard deviation of $\sim$ 0.4, we take this as a reasonable value for our random errors, of the same order as our predicted errors in $\phi$ and significantly smaller than the most conservative errors in $\phi$. We find the median value to be $\Xi_{\rm med} = 2.7$ cm$^{-3}$ pc. The results of this analysis are listed in Table \ref{param}.

\section{Density along the line of sight} \label{dls}

To determine the validity of these results we must compare them not only to existing distance estimates, but also to estimates for the density of the ISM along the line of sight to the HVCs. In the Galactic disk regime we simply assume a plane-parallel Cold Neutral Medium (CNM) and Warm Ionized Medium (WIM), as in \citet{DL1990} and \citet{Reynolds1993}. Since this region is observed at $b = -45^\circ$, once the plane-parallel assumption becomes questionable along the line of sight we will have left the disk regime and must concern ourselves only with the Hot Ionized Medium of the halo (HIM), which dominates the density.

In BD97 the assumption is made that the model for the halo at $R = R_\odot$ determined in W95 can be simply extrapolated as a plane-parallel medium, as we have for the other phases of the ISM. In our case, we cannot make this simple extrapolation for the halo, as our object lies at $R \sim 20$ kpc, where the character of the halo is rather different than it is locally. In a separate paper \citep{Peek2006}, the authors incorporate results from O{\textsc VII} absorption measurements \citep{fang2006}, halo emissivity measurements \citep{KS2000}, and recent theoretical constraints on disk pressure \citep{Wolfire2003} to develop a self-consistent, isothermal hydrostatic halo model. The model is essentially the simplest possible model of an isothermal gas in a known gravitational potential that is consistent with modern observational results. It is this global halo model to which we compare the H$\alpha$ distance and our dynamical results in Figure \ref{rhoD}. 

\begin{figure}
\includegraphics[scale=.34, angle=90]{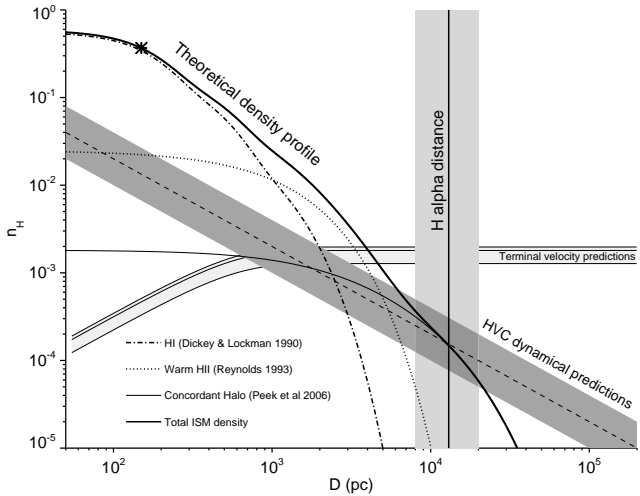}
\caption{This Figure is a plot of total ISM hydrogen number density, $\rm n_H$ vs. distance towards the HVC complex. The theoretical density profile is a sum of the neutral HI disk from \citet{DL1990}, the warm HII Reynolds layer from \cite{Reynolds1993} and the concordant halo model from Peek et al. 2006, in prep. The horizontal lines in the lightest shaded area are the predictions for the density of the ambient medium from the assumption of terminal velocity. The vertical line is the distance as predicted by H$\alpha$ measurements, with the associated shaded area representing error bars \citep{WVW2001}. Our dynamical estimate,  $\Xi_{\rm med}$ is plotted in a dashed line; the dark shaded region represents estimated errors of a factor of 2. The asterisk in the top left of the plot represents $\Xi_{\rm max}$ (see Equation \ref{Ximax}). Note that the intersection of the modeled $\rm n_H(D)$ with the measured distance is consistent with our result, and that the terminal velocity distance is not.}
\label{rhoD}
\end{figure}

\section{Discussion}

The results for $\Xi$ are in Table \ref{param}, along with the average and peak column density of each feature. These values vary by a factor of few, which is not surprising considering the assumptions with regard to morphological details of these clouds along the line of sight. In Figure \ref{rhoD}, we show the halo number density of hydrogen $n_H$ as a function of distance towards the cloud, as well as the H$\alpha$ distance from \citet{WVW2001}. We also plot the prediction, $\Xi$, from the dynamical analysis, along with the standard deviation within our sample. Our predictions are in good agreement with the intersection of the cumulative density profile and the H$\alpha$ distance.

In future observations, we can use our measured quantity, $\Xi$, to give us a handle on the distance to a cloud, if we assume the canonical density models in Figure \ref{rhoD}. Below about 2 kpc, where we take the halo to be dominated by the warm Reynolds Layer and the HI disk, the models show $n_H \propto D^{-3/2}$, so that $\Xi = n_H D \propto D^{-1/2}$  and our prediction for distance would be $D \propto \Xi^{-2}$. This implies that fractional errors in our quantity $\delta \Xi/\Xi$ will result in fractional errors in the distance ${\delta D}/{D} \propto 2 \delta \Xi/\Xi$. Evidently, our capacity to estimate distance is rather hampered in this nearby regime, as the errors that arise from confusion and projection effects are strongly amplified. This issue is of little relevance, though, as there are few direct measurements of HVCs so nearby (see \citei{Wakker2001}). The picture in the distant regime, 2 kpc $< D <$  100 kpc, is much rosier. Since in this region, $n_H \propto D^{-5/2}$, we have $D \propto \Xi^{-2/3}$, so that ${\delta D}/{D} \propto 2 \delta \Xi/3 \Xi$. This means that our errors in this figure lead to reduced errors in our distance measure, which gives a powerful measure of distance. Alternatively, for the few clouds that have good distance measurements, this technique could be turned around to measure densities and test theoretical models of the halo density profile and smoothness.

A concern with this kind of technique is that it could be applied to HVCs coincidentally along the same line of sight and thereby generate incorrect distance (or density) estimates. This is of course possible, but if the clouds have dissimilar velocities, $\Xi$ will increase dramatically. As $\Xi$ rises, the predicted distance decreases, but as the Galactic disk is approached the density profile, dominated by the HI disk, flattens out and the dynamical prediction no longer intersects it in any place. This defines the maximum possible value of $\Xi$: 

\begin{equation} \label{Ximax}
\Xi_{\rm max} = \frac{39 \unit{cm}^{-3} \unit{pc}}{{\rm cos} ~b},
\end{equation}
for the assumed \citet{DL1990} HI disk density profile (see asterisk in Figure \ref{rhoD}). Therefore, we can easily get an idea if it is impossible for two clouds to be related by drag forces - if their measured $\Xi$ is greater than this maximum value, then they cannot have been sheared apart by drag. 

We have shown that modeling the movement of various pieces of our HVC complex as having been stripped from the main body of the cloud by simple drag forces is consistent with reasonable halo models and the inferred H$\alpha$ distance. One exciting aspect of this conclusion is that it bolsters the  H$\alpha$ distance measure technique. Given both the large uncertainties regarding the magnitude and patchiness of the radiation field, and some uncertainty in whether the H$\alpha$ is generated solely by photoionization or whether collisional ionization plays a role, it is important that there exist checks on this distance measure. Since field patchiness can be stronger closer to the disk, our confirmation of the H$\alpha$ distance technique serves to show that patchiness is a limited effect at z $\simeq$10 kpc or higher, in agreement with \citet{B-HP2001}, at least toward this cloud. Collisional ionization is expected to increase with HVC velocity and halo density. Given the extreme velocities of these clouds, we claim that collisional ionization does not strongly effect the H$\alpha$ distance measurement for any HVCs in a similar density regime. 

We have inferred a distance above the plane for these clouds of $\sim 10$kpc. This calls into doubt whether these clouds could have been produced by a Galactic fountain model, where gas blow-out from the disk HIM launches material into the halo, which later recondenses and falls as HVCs (\citei{Bregman1980}, \citei{deAvillez2000}). Galactic fountain models set a maximum height for upflowing gas that condenses into HVCs at $\leq$10 kpc. Therefore any clouds at this height should be moving relatively slowly, as they are at the apex of their cycle, which contradicts the large velocities we measure. Also, at this distance, the complex has a total mass of $2.7 \times 10^5 M_{\odot}$ and a kinetic energy of $7.2 \times 10^{53}$ ergs. This is equivalent to the energy output of approximately 720 supernova explosions into the ISM. 

At our inferred position, W95 predicts a gravitational potential of $-1.16 \times 10^5 ($km s$^{-1})^2$. This implies that if the cloud started at rest far from the Galaxy and conserved energy as it fell, it would have a velocity of 480 km s$^{-1}$. As this velocity is in excess of the estimated space-velocity of the HVCs, we find that our distance is consistent with models in which HVC velocity stems from cloud free-fall in the Galactic potential. If this particular group of HVCs started from rest we can also determine that the HVCs could not have originated any closer than $\sim 60$ kpc.

The inferred density at this position is certainly in agreement with the observational constraints discussion in Section \ref{dls}, as they are used to construct the halo model to which we compare our results. The density is also in rough agreement with inferred halo densities, such as those put forward by  \citet{WW1996}, $n_h \ge 10^{-4}$cm$^-{3}$, and \citet{QM2001}, $n_h(10$ kpc$) <  3 \times10^{-3}$cm$^{-3}$.

Models in which HVCs form from the cooling of halo gas (e.g. \citei{MB2004}, \citei{Kaufmann06}, \citei{SL06}) allow comparison both in the HVC parameters and in the halo density. The HVC distance scale determined in \citet{Kaufmann06} of 10-20 kpc is in line with our observations and simulations in \citet{SL06} have cloud masses consistent with our derived cloud masses. \citet{SL06} also finds hot halo densities $10^{-3.5} \unit{cm}^{-3}> n_H > 10^{-4} \unit{cm}^{-3}$ at $R=20$ kpc,  and \citet{MB2004} find $n_H = 2 \times 10^{-4}$ at $R=20$ kpc in their ``hot gas core'', both in good agreement with our results. It seems that our measurements are in good agreement with this theoretical picture of HVCs and the halo. 

\section{Conclusion}

We have demonstrated that we see the effects of halo drag on a group of HVCs in their morphological and kinematic structure. Indeed, a simple model of drag physics gives a consistent and plausible result for the distance to these HVCs. This distance is in agreement with the H$\alpha$ distances measured for these clouds. At this distance, we confirm a halo density of $n_H \simeq 2 \times10^{-4} \unit{cm}^{-3}$ at $R \simeq 20$ kpc. These results are in rather good agreement with the modern theoretical picture of a circum-Galactic origin for HVCs, and in rather strong disagreement with typical Galactic fountain models.

To reproduce this analysis in other HVCs, one would need to both be able to identify similar morphological clues in the structure of HVCs (to determine $\Theta$) and be able to estimate $\phi$ with similar precision. While this is certainly possible in some other HVCs, we predict that the vast majority of HVCs will elude this technique in its current form. Whether the physics described herein influences the structure of larger HVC complexes and can be used to measure HVC distance and halo density has yet to be shown. In the few cases where this technique is usable, comparison to simulation would allow us to evaluate whether more sophisticated analyses are required to model fully the physics involved, or whether HVC distortion and destruction, for typical physical parameters, can be robustly modeled with our na\"ive gas-drag analysis. It is also interesting to note that the structure of a hydrostatic halo is a strong function of temperature on 10 kpc scales so any future measurements of $\Xi$ may help tightly constrain halo temperatures, and subsequently models of spiral galaxy formation and evolution. We look forward to more in-depth quantitative analysis of cloud and cloud-complex morphology in the presence of the Galactic halo.

The authors would like to acknowledge those who helped get GALFA going, including Jeff Mock, Aaron Parsons and Dan Werthimer as well as the Arecibo staff. JEGP would like to acknowledge many helpful conversations with Evan Levine and Leo Blitz. The research of CFM is supported in part by NSF grant AST00-98365 and NSF grant AST-060683. The research of CH and JEGP is supported in part by NSF grant AST04-06987.


\begin{center}
\begin{deluxetable*}{c c c c c c c c }
\tabletypesize{\scriptsize}
\tablecaption{Parameters for HVCs and features imaged. The parameters were determined by setting boundaries for the clouds where the error in the value for the height of the Gaussian fit to the spectra reached 25\%. The positions noted are the geometric centers of each of the features, and do not necessarily reflect the inferred values of $\Theta$. Features f-1, f-2 and f-3 were also bounded by a cut in velocity to distinguish them from the HVC-A. This cut was chosen at a minimum of the velocity histogram in the region.\label{param}}

\tablewidth{0pt}
\tablehead{
\colhead{Name} & \colhead{l} & \colhead{b} & \colhead{VGSR (km s$^{-1}$)} &  \colhead{$\langle{N_H}\rangle \left(10^{19}\rm{cm}^{-2}\right)$} & \colhead{${\rm N_{H,peak}\left(10^{19}\rm{cm}^{-2}\right)}$} & \colhead{$\Theta \left(^\circ\right)$} & \colhead{$\Xi \left(\rm{cm}^{-3} \rm{pc}\right)$} 
}
\startdata
HVC-A & 160.7 & $-44.8$ & $-280$ &  6.0 & 15 & --- &---  \\
HVC-B & 162.8 & $-46.3$ & $-257$ &  4.8 & 9.1 & 3 & 2.8\\ 
HVC-C & 156.3 & $-45.1$ & $-240$ &  2.8 & 4.1 & 4 & 2.3\\ 
HVC-U & 160.0 & $-48.4$ & $-236$ &  1.8 & 3.0  & 2 & 3.0\\
f-1 & 161.6 & $-44.9$ & $-270$ & 2.2 & 4.8 & 0.5 & 0.80 \\
f-2 & 160.7 & $-43.4$ & $-268$ & 2.5 & 7.2 & 0.4 & 1.8 \\ 
f-3 & 157.3 & $-46.6$ & $-267$ & 1.8 & 3.7 & 0.4 &1.3 \\ 
\enddata
\label{data}
\end{deluxetable*}
\end{center}

\end{document}